\documentclass[twocolumn,superscriptaddress,aps,pra]{revtex4-2}
\usepackage{bm}
\usepackage{amssymb,amsmath}
\usepackage{comment}
\usepackage{times}
\usepackage{braket} 
\usepackage[dvips]{graphicx}
\usepackage{array}
\usepackage[colorlinks,linkcolor=blue, urlcolor=blue, anchorcolor=blue, citecolor=blue]{hyperref}

\newcommand{\beq}{\begin{eqnarray}}
\newcommand{\eeq}{\end{eqnarray}}
\newcommand{\nn}{\nonumber}

\usepackage{ulem,lipsum}

\begin{document} 

\title{Scaling behavior of eigenspectrum for entanglement from correlation matrices}

\author{Lih-King Lim}
\email{lihking@zju.edu.cn}
\affiliation{Zhejiang Institute of Modern Physics, School of Physics, Zhejiang University, Hangzhou 310027, China}

\begin{abstract}
We study the scaling behavior of the eigenvalues of correlation matrices, which characterize the entanglement of a subsystem with its complement part of a total pure state. A most distinguishing feature of entanglement entropy is its logarithmic dependence on the subsystem size for the groundstate of one-dimensional critical systems. Despite its robust universal character and relevance to a wide range of topics, a thorough understanding of this result requires sophisticated mathematical physics techniques or conformal field theory. The aim of our work is to shed light on this from the underlying eigenvalue distribution perspective. The central object is the correlation matrix, which takes the form of Toeplitz or block-Toeplitz matrix. We develop a circulant matrix approximation in the large matrix dimension limit, thus allowing for the individual eigenvalues behavior to be analysed analytically. We find that for both free lattice fermions and transverse field Ising chain, eigenvalues in the bulk of the eigenspectrum scales as $1/L_A$ with the subsystem size. Together with the extensivity of the entropy function, it explains the robust $\log_2 L_A$ scaling of entanglement at criticality. Perturbing from the entanglement-free limit of the Ising chain, we find a smooth crossover behavior to `non-critical' scaling that is characterized by a very slow logarithmic dependence rendering it seemingly a constant entanglement value expected of non-critical systems.     
\end{abstract}

\date{\today}

\maketitle

\section{Introduction}
Entanglement is a most peculiar form of correlation in many-body quantum systems \cite{Amico08, Horodecki09, Eisert10, Dalmonte22}. Over the last decades, its investigations have found connections to many seemingly unrelated subjects, from quantum information \cite{Horodecki09} to strongly correlated systems \cite{Laflorencie16} and to cosmological black hole entropy \cite{Bombelli86}. In condensed matter systems, the entanglement structure of many-body quantum state offers new insights into quantum phase transitions \cite{Osterloh02, Vidal03}, long-time quantum dynamics \cite{Page93, Calabrese05, Popescu06, Bardarson12, Lazarides14,Beugeling15, Lim24, Lou25}, as well as novel many-body numerical techniques \cite{Schollwock05}. Quantum simulators form another frontier in studies of entanglement as many proposals are experimentally realized and studied \cite{Islam15, Joshi23, Tajik23, Google23, Chiu25}. In these context, the measure of bipartite entanglement of a pure state serves as the simplest and clearest diagnostic \cite{Bennett96}. 

In solvable models, models which are either non-interacting or mappable to free fermions, many rigorous results on entanglement have been established \cite{Page93, Peschel04, Plenio05, Vidmar17, Vidmar18}. This is thanks to the fact that all correlations of the system are described by correlation matrix of second moments, while higher moments are factorizable according to Wick's theorem \cite{Vidal03, Peschel03, Peschel09, Latorre04}. In particular, correlation matrix of a total pure state when restricted to a subpart (the subsystem) captures correlations coming solely from entanglement, namely the non-product-state nature of the two parts of a totally pure state. Hence the study of bipartite entanglement of a pure state becomes the study of correlation matrix for these solvable models entailing certain matrix structure. These models encompass non-interacting fermions hopping on a lattice as well as solvable spin chains such as the Ising and the XY models. However, they exclude general interacting spin models \cite{Swietek26} and interacting fermions \cite{Deutsch20}.

Correlation matrices, because of translation invariance of the system, take the form of the so-called Toeplitz or block-Toeplitz matrix, with matrix elements
\beq
C_{ij}=t_{j-i}, 
\eeq 
with $i,j=1,\ldots, L_A$, with $L_A$ being the size of the subsystem of interest. 
From their eigenvalues, one determines the reduced density matrix $\rho_A$ and evaluates to give the von Neumann entanglement entropy, $S_{vN}=-\textrm{Tr}_A [\rho_A \log_2 \rho_A]$, the canonical measure of bipartite entanglement. An important result from explicit analytical studies of the correlation matrix establishes that for critical ground state of one-dimensional system entanglement entropy exhibits a logarithmic dependence on the subsystem size with a universal coefficient $S_{vN}\!\!\sim\!\! (c/3) \log_2 L_A$ \cite{Vidal03, Korepin04, Jin04, Its09, Keating04}. This unique scaling behavior of entanglement entropy that interpolates an area and a volume law, respectively, of ground state and high-energy eigenstates of non-critical systems \cite{Vidmar18} has many ramifications, including the difficulty of density matrix renormalization group studies for critical states \cite{Schollwock05}.  The analysis, however, involves sophisticated mathematical results from Toeplitz matrices such as Fisher-Hartwig formula \cite{Jin04, Its09}. Most importantly, it establishes a direct connection to conformal field theory of critical systems identifying $c$ as the conformal charge \cite{Holzhey94, Vidal03, Korepin04, Calabrese04}, underscoring the universal appeal of the result. 

Indeed, in arriving at the result either by asymptotic analysis of Toeplitz matrix \cite{Jin04, Its09} or by conformal field theory \cite{Holzhey94, Calabrese04} is a technical feat of mathematical physics and field theory. Given the importance and its wide applicability, it is highly desirable to provide a simpler and more intuitive understanding of the result. Here we revisit the scaling behavior of entanglement entropy from the perspective of eigenspectrum of an asymptotically equivalent matrix in the form of circulant matrix. Perturbating matrix elements of the original Toeplitz matrix, on the one hand, can often lead to uncontrolled approximations and unphysical results,  for e.g., eigenvalues become non-probabilistic. Asymptotically equivalent circulant matrix \cite{footnote1}, on the other hand, generally preserves the underlying matrix structure and is shown (fulfilling certain mathematical conditions) to give the same functional value depending on the \textit{full} eigenspectrum in the large matrix dimension limit.

Circulant matrix is a special form of Toeplitz matrix with elements in the row related simply to the preceding row by a right cyclic shift.
Once constructed, it is much simpler to analyse as it allows for a straightforward analytical expression of its eigenvalues. 
The circulant approximation is actually widely studied in the field of signal processing and information theory, for e.g., to understand Szeg\"o's theorem on Toeplitz matrices \cite{Gray06}, but, to the best of our knowledge, much less used in statistical physics.  
Here we study the corresponding circulant matrix approximation for correlation matrices of free fermions and transverse field Ising model. 

We can show analytically that in the limit of large matrix dimension, the bulk of the eigenspectrum exhibits a $1/L_A$ behavior in both models at criticality. The logarithmic dependence of the entanglement entropy is a further result from the extensivity of the entropy function, see Fig.~\ref{fig:scaling}. The result therefore shows at the level of \textit{individual} eigenvalue scaling, something not evident from the Fisher-Hartwig formula, a result on the \textit{total} behavior of the eigenspectrum, i.e., study of Toeplitz determinant \cite{Jin04, Its09}. Furthermore, it is known that eigenvalues of the correlation matrix cluster closely to either 0 or 1, making an accurate evaluation of entropy a numerically demanding task, due to the embedded logarithmic function. So we do not expect the circulant approximation to reproduce exactly the original Toeplitz result, i.e., the rate of the logarithmic divergence differs. 

In addition, we also find that this behavior is not limited to critical systems as we extend our analysis to perturbation from the entanglement-free limit of the Ising chain. However a key overall scale factor can render a very weak logarithmic dependence in the latter. Our work thus offers new insights into the scaling bahavior of \textit{individual} eigenvalue of the correlation matrix \cite{footnote2, Franchini11} and the scaling of entanglement entropy from the asymptotically equivalent spectral perspective. 

The paper is organized as follows. In Sec.~\ref{sec:fermion} we introduce the single-band gapless free fermion model and the correlation matrix for the computation of the entanglement entropy. We then construct the circulant matrix approximation. The eigenvalues are computed analytically and we show our first result of critical eigenvalue scaling. In Sec.~\ref{sec:ising} we present computations for the transverse field Ising chain at the critical point. The correlation matrix in this case acquires a block-Toeplitz form, which requires a generalized circulant approximation. A more elaborated asymptotic analysis of the eigenspectrum then follows. In Sec.~\ref{sec:noncrit} we consider the non-critical case from perturbing the entanglement-free limit. In Sec.~\ref{sec:dis} we end with discussion and conclusions.

\section{Free fermions model and its entanglement entropy}
\label{sec:fermion}
\subsection{Free Fermions Model}
In this paper we study two models, one with non-interacting fermions, and the second with the transverse field Ising chain. The Hamiltonian of non-interacting spinless fermions hopping on a one-dimensional lattice takes the form of a tight-binding model:
\beq
H=-\sum_{<i,j>}\,a_i^\dag a_j+\textrm{h.c.}
\eeq
where $a_i^\dag$ ($a_i$) is the fermionic creation (annihilation) operator on site $i$. The total lattice sites is $L$ with fermion numbers $N_f$. The fermion density $n_f\equiv N_f/L$ is given by $0<n_f\leq1$. In this tight-binding limit with one atom per unit cell, it has a single energy band and therefore the ground state $| \Psi_0 \rangle$ with $n_f\leq1$ spinless fermions is always a gapless state, being `critical' in this sense. To study the groundstate entanglement entropy, we consider the correlation matrix \cite{Peschel09}:
\begin{figure}
\begin{center}
\includegraphics[width=7.7cm]{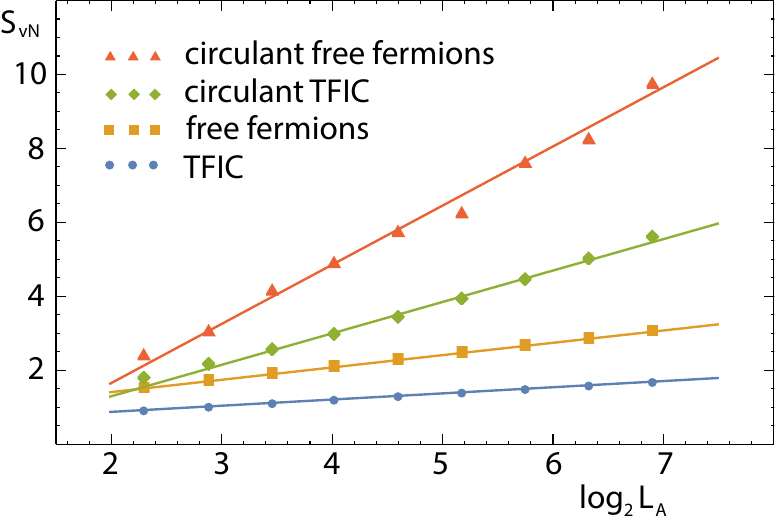}
\end{center}
\caption{Scaling behavior of entanglement entropy $\sim (c/3)\log_2 L_A$ for free lattice fermions (squares) and transverse field Ising chain (circles) at criticality. They obey conformal field theory result with $c=1$ and $c=1/2$, respectively.  The circulant approximation to the original problems, represented with triangles and diamonds, respectively, giving the logarithmic dependences.}\label{fig:scaling}
\end{figure}
\beq
[\mathbf{C}_A]_{ij}=\langle \Psi_0 |a_i^\dag a_j| \Psi_0 \rangle
\eeq
where the lattice index $i,j\in A$ is restricted to $L_A$ contiguous sites. The non-interacting nature of the system allows for its entanglement characterization via $\mathbf{C}_A= U^\dag \, \textrm{diag} (p_1,p_2,\ldots,p_{L_A})\, U$ after a unitary transformation, with its eigenvalues $p_i$ being the occupation probabilities of the effective free modes of subsystem A with the reduced density matrix
\beq
\rho_A &=& \left( \begin{array}{cc} p_1 & 0 \\ 0 & 1-p_1 \end{array} \right)\otimes\left( \begin{array}{cc} p_2 & 0 \\ 0 & 1-p_2 \end{array} \right) \nn\\
&& \otimes \cdots \otimes \left( \begin{array}{cc} p_{L_A} & 0 \\ 0 & 1-p_{L_A} \end{array} \right),
\eeq
see Fig.~\ref{fig:eigen1}(a). Note that while the state of the total system is a pure state, it is only after tracing out the complement degree of freedom that we obtain a mixed state description for the subsystem density matrix $\rho_A$ due to entanglement between the two parts. The entanglement entropy of subsystem A is then given by
\beq
S_{vN}&=&-\textrm{Tr}_A[\rho_A \ln \rho_A ]\nn\\
&=& - \sum_{i=1}^{L_A}\,\bigl[p_i \ln p_i +(1-p_i) \ln (1-p_i)\bigr]. 
\eeq
\subsection{Correlation Matrix and Circulant Approximation}
We now turn to the correlation matrix element for the groundstate of the free fermions chain given by \cite{Peschel09}
\beq
\langle a_i^\dag a_j \rangle &=& \frac{1}{L}\frac{\sin(n_f \,\pi (i-j) )}{\sin(\frac{\pi}{L} (i-j))}\nn\\
&\rightarrow& \frac{\sin((n_f\, \pi (i-j) )}{\pi (i-j)} \,\,\,\,\textrm{for large-$L$.} 
\eeq
This renders the correlation matrix $\mathbf{C}_A$ to have the $L_A\times L_A$ Toeplitz form 
\beq
\left( \begin{array}{cccccc}
t_0 & t_{-1} & t_{-2} &\ldots & \ldots & t_{-(L_A-1)} \\
t_1 & t_0 & t_{-1} & t_{-2} & \ldots & t_{-(L_A-2)}\\
\vdots & \vdots & \ddots &  & & \vdots\\
t_{L_A-2} & t_{L_A-3} & \ldots & t_1 & t_0 & t_{-1}\\  
t_{L_A-1} & t_{L_A-2} & & \ldots & t_1 & t_0
\end{array} \right)
\eeq
where $t_{i-j}=\langle a_i^\dag a_j \rangle=t_{j-i}$. We are interested in the large matrix dimension limit, that is the behavior of the eigenvalues as the subsystem size $L_A$ grows large. We will now construct a tractable circulant matrix approximation to the original Toeplitz problem with \cite{Gray06, Pearl73, Zhu17}
\begin{figure}
\begin{center}
\includegraphics[width=7.7cm]{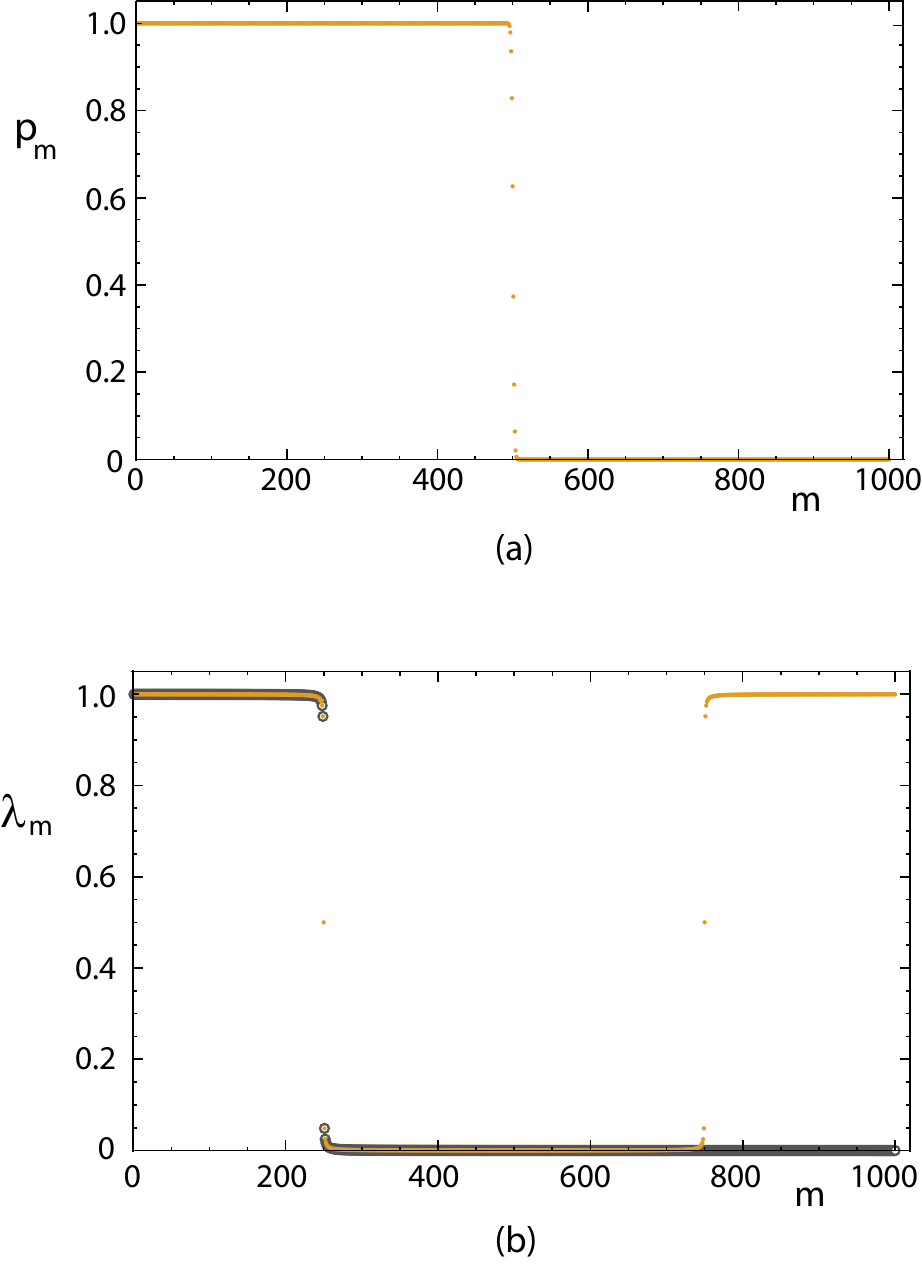}
\end{center}
\caption{(a) Exact eigenvalue distribution from the correlation funtion for free lattice fermions. (b)  Eigenvalue distribution from the circulant matrix approximation. Full dots (in yellow) are from the discrete summation. Open circles (in grey) are in the continuum limit. Here $L_A=1000$, $n_f=0.5$.}\label{fig:eigen1}
\end{figure}
\beq\label{cir}
\left( \begin{array}{cccccc}
c_0 & c_{1} & c_{2} &\ldots & \ldots & c_{L_A-1} \\
c_{L_A-1} & c_0 & c_{1} & c_{2} & \ldots & c_{L_A-2}\\
\vdots & \vdots & \ddots &  & & \vdots\\
c_{2} & c_{3} & \ldots & c_{L_A-1} & c_0 & c_{1}\\  
c_{1} & c_{2} & & \ldots & c_{L_A-1} & c_0
\end{array} \right)
\eeq 
with the matrix element
\beq\label{cmt}
c_j=   \frac{L_A-j}{L_A}\, t_{-j}+ \frac{j}{L_A}\, t_{L_A-j}.   
\eeq
This circulant matrix has been shown to be asymptotically equivalent to the original Toeplitz matrix, in that the total behavior of the two sets of eigenvalues are equivalent in the large-$L_A$ limit \cite{Gray06, Pearl73, Zhu17}. This allow us to address the Toeplitz matrix problem with a much simpler circulant matrix, and we do not need the mathematically sophisticated asymptotic analysis with Fisher-Hartwig formula \cite{Its09, Bottcher99, Puel23}.

Specifically, the $m$-th eigenvalue of Eq.~(\ref{cir}) is given by 
\beq\label{ei1}
\lambda_m = \sum_{j=0}^{L_A-1}\,c_j e^{i \frac{2 \pi}{L_A}j \cdot m },\,\,\,\,m=1,2,\ldots,L_A,
\eeq
and the associated eigenvector
\beq
(1,e^{i \frac{2 \pi}{L_A} m}, e^{2 i \frac{2 \pi}{L_A} m}, e^{3 i \frac{2 \pi}{L_A} m},\ldots,e^{(L_A-1) i \frac{2 \pi}{L_A} m})^T,
\eeq
defining the eigenvalue problem with circulant matrices. In the large-$L_A$ limit, we turn the discrete summation of Eq.~(\ref{ei1}) into integral introducing the variable $k=2\pi j /L_A$ to give
\beq\label{ei2}
\lambda_m =  \int_{0}^{2\pi} \frac{dk}{2\pi}\,  c_k \,e^{i k m}
\eeq
where 
\beq
c_k &=&  \frac{2\sin(n_f L_A k/2)}{k}+\frac{k \sin(n_f L_A (2\pi-k)/2)}{2\pi-k}\nn\\
&&-\frac{1}{\pi}\sin(n_f L_A k/2).
\eeq
By performing the Fourier transforms we obtain the eigenvalues expression
\beq
\lambda_m&=&\frac{1}{\pi}\, \textrm{Si}\left( 2\pi\, L_A \left(\frac{1}{2}n_f-\frac{m}{L_A}\right) \right)\nn\\
&&+\frac{1}{\pi} \textrm{Si}\left( 2\pi\, L_A \left(\frac{1}{2}n_f+\frac{m}{L_A}\right) \right)\nn\\
&&+\frac{1}{\pi^2 L_A}\, \frac{(n_f/2) \sin^2(\pi n_f L_A/2 )}{(m/L_A)^2-(n_f/2)^2}, 
\eeq
with $m=1,2,\ldots,L_A$,  see Fig.~\ref{fig:eigen1}(b),  and the sine integral function is defined as $\textrm{Si}(x)\equiv\int_0^x \frac{\sin t}{t} dt$.  

With this expression, we can perform the scaling analysis. There are two parameters in the problem, namely, $L_A$  and $m$. The latter runs from $1$ to $L_A$ labelling the eigenvalue of the $L_A\times L_A$ circulant matrix. So, the large-$L_A$ limit is obtained by letting $L_A$ large \textit{and} keeping $m/L_A$ constant. The fermion density $n_f$, ranging from $0$ to $1$, is another constant to be kept fixed. 

Using the asymptotic expression for sine integral function  
\beq
\textrm{Si}(x)\sim \pm \frac{\pi}{2} \mp \frac{\cos(x)}{|x|}+\mathcal{O}(1/x^2), \textrm{\ \ for $x\rightarrow \pm\infty$,}
\eeq  
with a leading correction as damped oscillating behavior with amplitude decaying as $1/|x|$. The function behaves like a smoothed Heaviside step-function interpolating the values $\pm \pi/2$ at the transition point $x=0$. 

For large $L_A$ and consider first $m/L_A < n_f/2$ we get
\beq
\lambda_m&\approx&1-\frac{1}{\pi^2 L_A} \, \frac{\frac{1}{2}n_f }{(n_f/2)^2-(m/L_A)^2}\cos^2 \left(\pi \frac{n_f}{2} L_A\right)   \nn\\
&\equiv&  1-\frac{a'}{L_A}.
\eeq
Similarly, for large $L_A$ and consider $m/L_A > n_f/2$ we then have 
\beq
\lambda_m&\approx&\frac{1}{\pi^2 L_A} \,\frac{\frac{1}{2}n_f }{(m/L_A)^2-(n_f/2)^2}\cos^2 \left(\pi \frac{n_f}{2} L_A\right)    \nn\\
&\equiv& \frac{b'}{L_A}
\eeq 
We introduce the positive prefactor constants $a'$ and $b'$ that maintain the same order of magnitude as $m/L_A$ is varied in the two respective ranges, either $m/L_A<n_f/2$ or $m/L_A>n_f/2$, at fixed $L_A$ and $n_f$. Moroever, besides in $m/L_A$, $L_A$ enters only in the argument of the cosine function squared and it does not change the overall $1/L_A$ behavior. 

We find that the eigenvalues behave as expected like occupation probabilities of the effective free modes labelled by $m$, see Fig.~\ref{fig:eigen1}. Most of the eigenvalues takes value either close to $0$ or $1$ with the transition point in the range $m/L_A\approx n_f/2$. 

We can now address the bulk scaling bahavior of the entanglement entropy by considering the scaling behavior of the individual eigenvalues. In the entanglement entropy function for each term, substituting the asymptotic values of the eigenvalue $\lambda_m\approx 1-\delta$ or $\delta$ with $\delta\equiv(a',b')/L_A$, we get 
\beq
&&-\lambda_m \ln \lambda_m-(1-\lambda_m) \ln (1-\lambda_m)\nn\\
&&\,\,\,\,\,\,\,\,\,\,\,\,\,\,\,\approx  -\delta \ln \delta + \mathcal{O}(\delta).
\eeq
The entanglement entropy is the sum of roughly $L_A$ terms of this form, except for the regime close to the transition point $m \approx \frac{1}{2}n_f L_A$. The sum then gives a bulk contribution of roughly $L_A$ terms of the order $-\delta \ln \delta$, thus giving the scaling behavior
\beq
S_{vN} \approx - L_A \, \delta \ln \delta  \sim \ln L_A.
\eeq
Fig. \ref{fig:scaling} shows the scaling of entanglement with the full circulant approximation (in triangles), agreeing with the qualitative argument given above. However, the slope differs from the microscopic or conformal field theory's universal coefficient. 

\section{Ising chain and Entanglement Entropy Scaling}\label{sec:ising}
\subsection{Tranverse Field Ising Chain and its Correlation Matrix}
The Hamiltonian is
\beq
H=-\frac{1}{2}\sum_{j} [\sigma_j^{x}\sigma_{j+1}^{x}+h\sigma_{j}^z]
\eeq
where $\vec{\sigma}_i$ is the Pauli matrix operating at the site $i$, $h$ is the external magnetic field. The subsystem is made of $L_A$ contiguous spins in an infinite chain. The system is at the quantum critical point $h=1$, which is the main focus of our work. To study the entanglement entropy of the groundstate, it is solvable with correlation function method (for $h\geq 1$), much similar to the free fermions model except that the matrix acquires a block-Toeplitz form. The correlation matrix is given by \cite{Vidal03, Calabrese05, Yang74}
\beq
\Gamma_{L_A}=\left( \begin{array}{cccc}
\hat{\Pi}_0 & \hat{\Pi}_{-1} & \ldots & \hat{\Pi}_{-(L_A-1)} \\
\hat{\Pi}_{1} & \hat{\Pi}_0 &  & \vdots \\
\vdots &  & \ddots & \vdots  \\
\hat{\Pi}_{L_A-1} & \ldots & \ldots & \hat{\Pi}_0 
\end{array} \right),
\eeq
with
\beq
\hat{\Pi}_j=\left( \begin{array}{cc}
0&g_j\\
-g_{-j}&0 \end{array} \right),
\eeq
and the matrix element
\beq\label{angle}
g_j&=&\int_0^{2\pi}  \frac{d k }{2\pi} \,e^{-i j k}\,\frac{\cos k-h-i\sin k}{|\cos k -h-i\sin k|}\nn\\
&\overset{\text{$h=1$}}{=}&\int_0^{2\pi}  \frac{d k}{2\pi} \,e^{-i j k}\, e^{i(3\pi/2- k/2)}=\frac{(-1/\pi)}{ \frac{1}{2}+j} ,
\eeq
where in the second line we focus specifically at the critical point $h=1$. Moreover, we have taken the total system size $L\rightarrow \infty$. 

\begin{figure}
\begin{center}
\includegraphics[width=7.7cm]{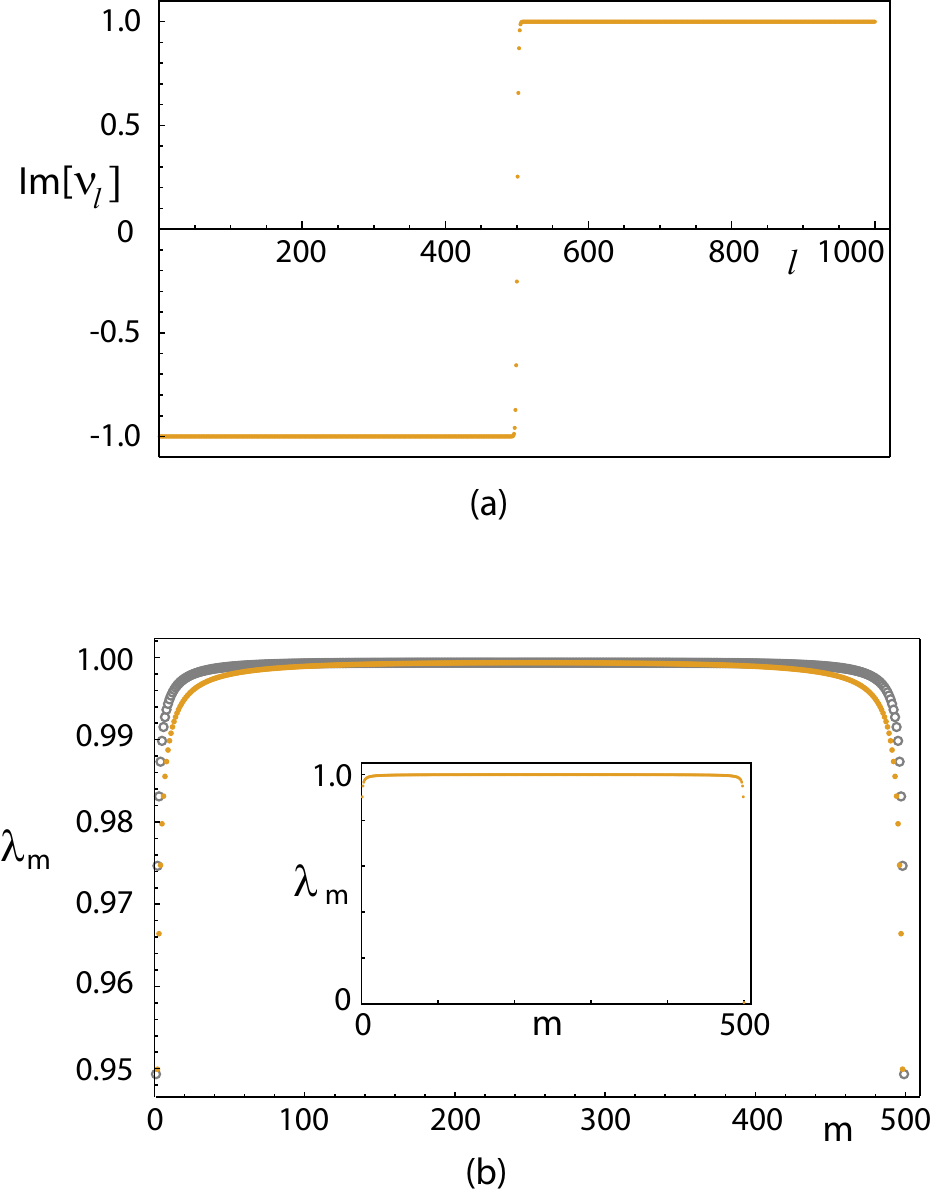}
\end{center}
\caption{(a) Exact eigenvalue distribution (purely imaginary) from the correlation matrix for the transverse field Ising chain at the quantum critical point, with $l=1,2,\ldots, 2 L_A$. (b) Comparison of the circulant matrix eigenvalues Eq. (\ref{trans1}) (full dots in yellow), with the large-$L_A$ expression Eq. (\ref{trans2}) (open circles in grey). Inset: The full eigenvalue distribution with the circulant approximation. Here $L_A=500$.}\label{fig:eigen2}
\end{figure}

The eigenvalues of $\Gamma_{L_A}$ come in pairs as $\pm i\nu_{m}$, $m=1,\ldots,L_A$, with $\nu_m$ bounded between $-1$ and $+1$, see Fig.~\ref{fig:eigen2}(a). The entanglement entropy is given by $S_{vN}=\sum_{m=1}^{L_A}\,H(\nu_m)$, where $H(x)$ is
\beq\label{ee2}
H(x)=-\frac{1+x}{2}\ln\frac{1+x}{2}-\frac{1-x}{2}\ln \frac{1-x}{2}.
\eeq

\subsection{Circulant Approximations}
We generalize the circulant approximation method \cite{Gray06, Pearl73, Zhu17} to the block-circulant form:
\beq\label{cir2}
\left( \begin{array}{cccccc}
\hat{c}_0 & \hat{c}_{1} & \hat{c}_{2} &\ldots & \ldots & \hat{c}_{L_A-1} \\
\hat{c}_{L_A-1} & \hat{c}_0 & \hat{c}_{1} & \hat{c}_{2} & \ldots & \hat{c}_{L_A-2}\\
\vdots & \vdots & \ddots &  & & \vdots\\
\hat{c}_{2} & \hat{c}_{3} & \ldots & \hat{c}_{L_A-1} & \hat{c}_0 & \hat{c}_{1}\\  
\hat{c}_{1} & \hat{c}_{2} & & \ldots & \hat{c}_{L_A-1} & \hat{c}_0
\end{array} \right)
\eeq 
with the block-matrix element
\beq\label{cmt2}
\hat{c}_j=   \frac{L_A-j}{L_A}\, \hat{\Pi}_{-j}+ \frac{j}{L_A}\, \hat{\Pi}_{L_A-j}.   
\eeq
Despite the block-Toeplitz and block-circulant forms, in the limit of large $L_A$, the total eigenvalue behavior of $\Gamma_{L_A}$ and Eq.~(\ref{cir2}) can be numerically shown to be equivalent, just like the simple Toeplitz and its circulant form.

To solve the eigenvalue problem of Eq.~(\ref{cir2}) we introduce the 2-block eigenvector $\Psi^T=(a,b)$ to give 
\beq
\sum_{j=0}^{L_A-1} \hat{c}_j e^{i \frac{2\pi}{L_A}j\cdot m} \Psi_m = \lambda_m \Psi_m,
\eeq
with $m=1,2,\ldots,L_A.$ Explicitly we have to solve 
\beq\label{gm}
\left( \begin{array}{cc}
0& G_m\\
-F_m &0
\end{array} \right)  \left( \begin{array}{c}
a_m\\
b_m
\end{array} \right)=\lambda_m  \left( \begin{array}{c}
a_m\\
b_m
\end{array} \right)
\eeq
with
\beq
&&G_m=\frac{1}{L_A} \sum_{j=0}^{L_A-1}    \,(\,(L_A-j)\,g_{-j}+j\, g_{L_A-j}\,) e^{i \frac{2\pi}{L_A}j\cdot m}\nn\\
&&F_m= \frac{1}{L_A} \sum_{j=0}^{L_A-1}    \,(\,(L_A-j)\,g_{j}+j\, g_{-(L_A-j)}\,) e^{i \frac{2\pi}{L_A}j\cdot m} \nn\\
&&\,\,\,\,\,\,\,\,\,\,= G_m^*.
\eeq
In the second equation we shifted the variable $j=L_A-j'$, the exponential factor being periodic in $j$, and the two terms $j'=0$ and $j'=L_A$ are identical. Eq.~(\ref{gm}) can be decoupled to give $-G_m G^*_m\, a_m = \lambda_m^2 \,a_m$ and $ -G_m^* G_m\, b_m = \lambda_m^2 \,b_m$. Substituting the matrix elements $g_j=(-1/\pi)/(1/2+j)$ we obtain the eiganvalues (i.e., writing $G^*_m$ part)
\beq\label{gm2}
\lambda_m =\pm \frac{i}{\pi L_A} \biggl\vert \sum_{j=0}^{L_A-1} \left( \frac{L_A-j}{\frac{1}{2}+j} + \frac{j}{\frac{1}{2}-L_A+j}\right) e^{i \frac{2 \pi}{L_A} j\cdot m}\,\biggr\vert .
\eeq

It turns out that the series in Eq. (\ref{gm2}) is slowly converging and cannot be calculated by integration in the continuum limit $L_A\rightarrow \infty$. However, the series can be manipulated exactly with the use of transcendental functions giving the analytical results: 
\beq\label{trans1}
\lambda_m&=&\pm\frac{i}{\pi}\,\biggl\vert\,-\frac{1}{L_A}+4 \, e^{-i\frac{m \pi}{L_A}} \,\textrm{arctanh}\left( e^{i\frac{m \pi}{L_A}}\right) \nn\\
&&-\,(1+\frac{1}{2 L_A})  \,\Phi\left(e^{i\frac{2 m \pi}{L_A}},\,1,\,1/2+L_A \right)\nn\\
&&-(1-\frac{1}{2 L_A}) \,e^{i\frac{2 m \pi}{L_A}}\, \Phi\left(e^{i\frac{2 m \pi}{L_A}},\,1,\,3/2-L_A \right)\biggr\vert .
\eeq
The Lerch transcendent is defined as $\Phi(z,s,a)\equiv \sum_{k=0}^{\infty}\,\frac{z^k}{(k+a)^s}$ \cite{Wiki:Lerch}. These expressions can be numerically verified to be exact comparing the discrete sum Eq.~(\ref{gm2}) and the one using transcendental functions Eq.~(\ref{trans1}).

\subsection{Scaling Analysis of Eigenvalues}
We now analyse the scaling behavior of the eigenvalues in the large $L_A$ limit. Note that the large $L_A$ limit is taken along with keeping $m/L_A$ a finite constant. We write the second term as $\textrm{arctanh}\left( e^{i\frac{m \pi}{L_A}}\right)=i\,\frac{\pi}{4} +\frac{1}{2} \log\left( \cot\left(\frac{m\pi}{2L_A}\right) \right)$ with $\textrm{arctanh}(z)=\frac{1}{2}\ln \left(\frac{1+z}{1-z}\right)$. In the large $L_A$ limit the last two factors scale, respectively, as 
\begin{widetext}
\beq
\Phi\left( e^{i\frac{2 m \pi}{L_A}},1,\frac{1}{2}+L_A \right)&\sim&\frac{1}{2L_A}+i\,\frac{1}{2 L_A}\,\cot\left( \frac{ m \pi}{L_A}\right)+\mathcal{O} \left(\frac{1}{L_A^2}\right),\nn\\
e^{i\frac{2 m \pi}{L_A}}\Phi\left( e^{i\frac{2 m \pi}{L_A}},1,\frac{3}{2}-L_A \right)&\approx& 2\, e^{-i \frac{m\pi}{L_A}} \,\log\left( \cot\left(\frac{m\pi}{2 L_A}\right) \right)-\frac{1}{2 L_A}+\mathcal{O} \left(\frac{1}{L_A^2}\right).
\eeq  
\end{widetext}
The first expression is obtained from an asymptotic expansion of the Lerch transcendent for large $L_A$ and constant $m/L_A$: $\Phi(z,1,a)\sim \frac{1}{1-z}\frac{1}{a}+\mathcal{O}(1/a^2)$ for large positive $a$ \cite{Wiki:Lerch}. The second expression, with large negative $a$, there is no asymptotic expression for it. We determine numerically for large $L_A$ and constant $m/L_A$. The first constant term is determined from the cancellation of the part from the arctanh term, followed by a $1/L_A$ correction. We finally obtain the large-$L_A$ expression for the eigenvalue: 
\beq\label{trans2}
\lambda_m&\approx&\pm i\, \left(1-\frac{1}{2\pi L_A}\,\frac{1+\sin^2\left(\frac{m\pi}{L_A}\right)}{\sin\left(\frac{m\pi}{L_A}\right)}\right)\nn\\
&\equiv&\pm i\, \left(1-\frac{c'}{L_A}\right),
\eeq
see Fig. \ref{fig:eigen2}(b). The eigenvalues at the the critical point behave exactly like the ones we studied in the single-band free fermion case. Plugging into Eq. (\ref{ee2}) the summation term for the entropy function leads to
\beq
H(1-\frac{c'}{L_A})\approx -\frac{c'}{2 L_A} \ln \frac{c'}{2 L_A} 
\eeq 
that is weakly dependent on $m/L_A$ for $m/L_A$ away from $0$ and $1$. So the total summation of roughly $L_A$ terms gives a final scaling relation $S_{vN}\sim \ln L_A$ at the critical point, see Fig. \ref{fig:scaling} (in diamonds).
\begin{figure}
\begin{center}
\includegraphics[width=6.5cm]{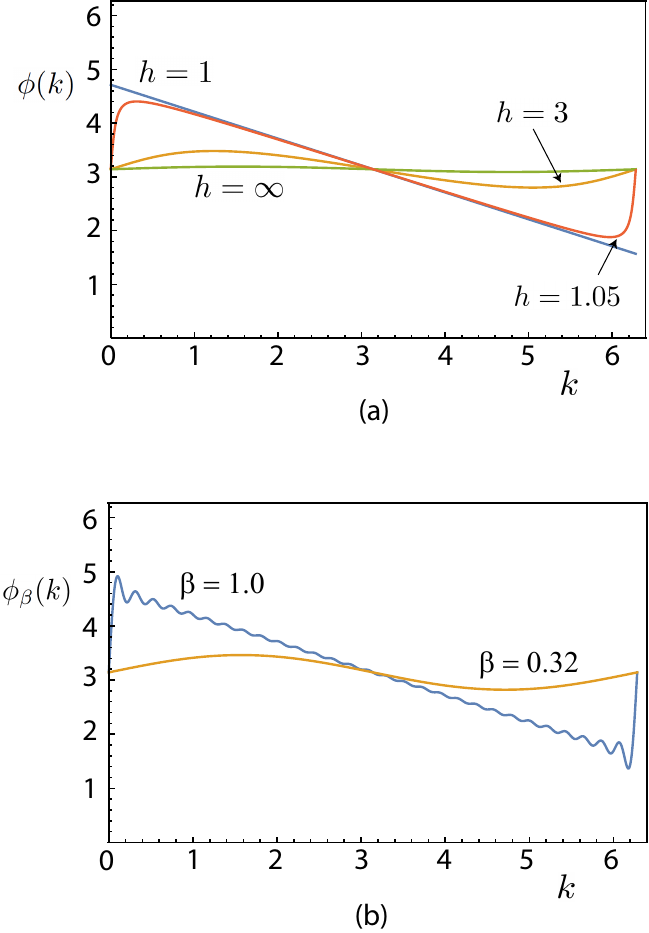}
\end{center}
\caption{(a) The phase profile $\phi(k)$ of the transverse field Ising chain for various magnetic field values: $h=1, 1.05, 3$ and $h=\infty$. (b) The sawtooth function Eq. (\ref{sawt}) with $\beta=1$ containing 30 harmonics (in blue) to approximate the phase profile for $h\rightarrow 1$, and $\beta=0.32$ with first harmonics for $h\approx 3$ (in yellow).}\label{fig:phase}
\end{figure}

\section{Non-Critical case}\label{sec:noncrit}
Tuning away from the critical point, we expect the scaling behavior gradually deviates from the critical behavior. The intuition is as follows \cite{Vidal03}. The distance to the critical point, the ``correlation length" $\xi$, sets a scale in comparison to the subsystem length of interest for examining the entanglement scaling. When $\xi$ is larger than the subsystem length the entanglement behavior resembles that of a critical system. As the subsystem length increases beyond $\xi$ then the scaling behavior changes to the non-critical behavior. To describe the complete crossover is quite challenging. Next we study the non-critical Ising chain starting from the entanglement-free-limit at $h=\infty$. 

\subsection{Ising chain at $h=\infty$}
Referring to the correlation matrix elements, from Eq. (\ref{angle}), $g_j=\int_0^{2\pi}  \frac{d k}{2\pi} \,e^{-i j k}\,e^{i \phi(k)}$, the behavior of $g_j$ as a function of separation $j$ is determined by the phase profile $\phi(k)$ as a function of $k$.  At the critical point $h=1$, the phase is $\phi(k)=3\pi/2-k/2$ that interpolates linearly between the values $3\pi/2$ and $\pi/2$. On the other hand, in the infinite field limit $h\rightarrow \infty$, the phase becomes a constant $\phi(k)=\pi$, see Fig.~\ref{fig:phase}(a). The matrix elements then take the simple form $g_j=-\delta_{j,0}$, giving eigenvalues of $\pm i$ hence a zero entanglement entropy state, i.e., in the infinite field limit all spins are aligned with the external field, a product state. Thus, the phase exhibits a highly non-perturbative behavior in deviating from the critical point.

\begin{figure}
\begin{center}
\includegraphics[width=7.8cm]{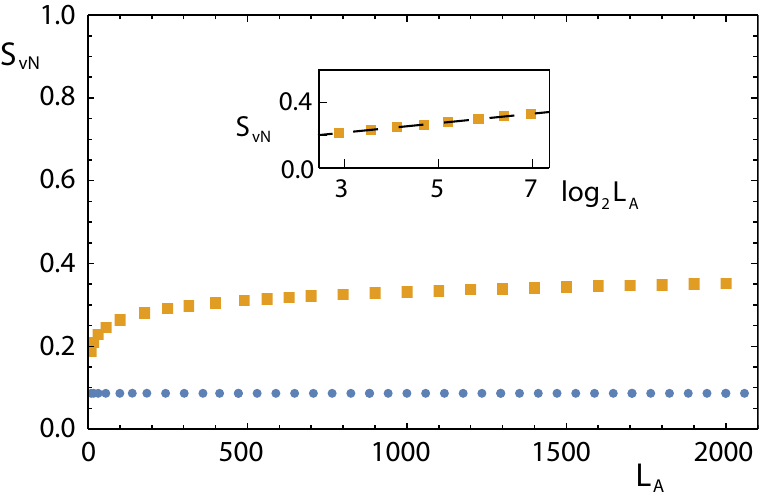}
\end{center}
\caption{Scaling behavior of entanglement entropy for non-critical transverse field Ising chain from exact correlation function (circles) and circulant matrix (squares) for $h=3$. Inset: Linear-log plot for the entanglement from the circulant matrix showing a very weak logarithmic dependence on the subsystem size fitted with $ (1/33) \log_2 L_A + 0.12$.}\label{noncrit}
\end{figure}

To describe the crossover from the critical case to the zero entanglement phase, we extend the momentum range of $\phi(k)$ to a repeated zone scheme, so that its critical field limit takes the form of a sawtooth function with infinitely many harmonics:
\beq\label{sawt}
\phi_\beta(k)=\pi+\beta\left[   \sin (k)+\frac{1}{2}\sin (2k) +\frac{1}{3}\sin (3k)+\cdots           \right],
\eeq
with $\beta\approx 1$, whereas in the infinite field limit, the phase takes the constant value $\pi$ with $\beta\rightarrow0$. In the large field limit, we then approximate the phase with the first harmonic correction $\phi_\beta(k)=\pi+\beta \sin(k)$, which, for e.g., by taking $\beta\approx 0.32$ it reproduces the field value of $h\approx 3$, Fig.~\ref{fig:phase}(b). In this case the matrix elements are given by
\beq
g_{\pm j}&=& - J_{|j|}(\beta)\,\,\,\,\,\,\textrm{for }\,\,\,\,j=0,2,4,\ldots;\nn\\
g_{\pm j}&=& \mp J_{|j|}(\beta)\,\,\,\,\,\,\textrm{for }\,\,\,\,j=1,3,5,\ldots,
\eeq
where $J_l(x)$ is the Bessel function of the first kind. From Eq. (\ref{gm}) we get for the eigenvalue $-\lambda_m^2=|G_m|^2$ with
\beq
G_m&=&-J_0(\beta)-\frac{L_A-2}{L_A}\,J_2(\beta)\, 2 \cos(4 \pi m/L_A)+\ldots\nn\\
&&+i\left[  \frac{L_A-1}{L_A}\,J_1(\beta) \,2 \sin(2 \pi m/L_A) +\ldots\right]
\eeq with $J_j(\beta)$ a rapidly decaying function of $j$, and so, we keep only up to $l=2$ in the expression. Furthermore we make use of the small argument expansion of the Bessel function up to the second order in $\beta$:
\beq
J_0(\beta)=1-\frac{\beta^2}{4},\,\,\,J_1(\beta)=\frac{\beta}{2},\,\,\,J_2(\beta)=\frac{\beta^2}{8}.
\eeq
Substituting into the eigenvalue equation we get
\beq
\lambda_m
&\approx&\pm i \biggl(1-\frac{\beta^2}{2}+\frac{L_A-2}{L_A}\frac{\beta^2}{2}\cos(4\pi m/L_A)\nn\\
&&+(\frac{L_A-1}{L_A})^2 \beta^2\sin^2(2\pi m/L_A)\biggr)^{1/2}\nn\\
&\approx&\pm i \left(1-\frac{\beta^2}{2 L_A}\right)
\eeq
for $m=1,2,\ldots, L_A$. In this case, the eigenvalues are constants independent of $m$, but continue to show the scaling behavior of $1/L_A$, similar to the critical cases studied in the last two sections.

Following the same argument with bulk contributions to the entanglement entropy, we get the leading scaling behavior of $\sim\beta^2 \log_2 L_A$. We see that the entanglement remains a logaritmic function on the subsystem size, despite being in a non-critical case. However, it depends on an overall scale factor $\beta^2$, being a perturbation parameter, crossing over to the entanglement free limit $S_{vN}\rightarrow 0$ as $h\rightarrow \infty$ and $\beta \rightarrow 0$. Since $\beta$ is a small parameter, the dependence on the subsystem length is therefore a very weak logaritmic dependence, see Fig.~\ref{noncrit} (note the scale).  

\section{Discussion and Conclusions}\label{sec:dis}
In this paper we provided a spectral perspective for understanding both critical and non-critical scalings of entanglement entropy of one-dimensional free fermions in the lattice and the transverse field Ising chain. We analytically obtained the scaling behavior of \textit{individual} eigenvalues of asymptotically equivalent circulant matrices from the original Toeplitz problems, showing a robust $1/L_A$ scaling in the bulk of the eigenspectrum. Together with the extensivity of the entropy function, it then led to the logarithmic dependence of the entanglement entropy. The rate of divergence, however, differs from the microscopic consideration, which can be due to the accuracy of the asymptotic equivalence of the two sets of matrices. Specifically, the entropy function in the form $x \log_2 x$ being non-analytic at origin may require a sharper condition than asymptotic equivalence. Other scheme of circulant approximation can be envisaged in future work. Finally, the method we developed here can be applied to other related Toeplitz problems, such as studying mid-spectrum eigenstates entanglement \cite{Vidmar18, Tian21}, and Toeplitz matrices with random elements \cite{Bogo20, Bogo21}.   

\begin{acknowledgments}
We thank Z. Liu, P. Ribeiro, C. Tian, and X. Wan for helpful discussions. 
This work is supported by the Zhejiang Provincial Applied Basic Research Program (Grant No. 2026C02A2005). 
\end{acknowledgments}


\begin{thebibliography}{99}
\bibitem{footnote1}
Two $n\times n$ matrices, $A_n$ and $B_n$, are said to be asymptotically equivalent when $\lim_{n\rightarrow \infty} |A_n-B_n|=0$, where $|O_n|\equiv (\frac{1}{n}\textrm{Tr} [O_n^\dag O_n])^{1/2}$ is the Hilbert-Schmidt norm for matrix $O_n$. For more details, see e.g., Ref.~\cite{Gray06}.
\bibitem{footnote2}
Corner transfer matrix method has been developed to study the spectral property of the reduced density matrix \cite{Peschel09}. It only works for the case of spin chains cut in the middle, i.e., $L_A/L=1/2$ is kept fixed. This limit is different from the limit studied in our work and others by first letting $L\rightarrow \infty$ and study the large-$L_A$ bahavior.
\end{thebibliography}
\end{document}